# EPR of photochromic $Mo^{3+}$ in $SrTiO_3$

Th. W. Kool

Van 't Hoff Institute for Molecular Sciences, University of Amsterdam
NL 1018 WV Amsterdam, the Netherlands
*March 2010*

*Abstract*

In single crystals of $SrTiO_3$, a paramagnetic center, characterized by $S = 3/2$ and hyperfine interaction with an $I = 5/2$ nuclear spin has been observed in the temperature range 4.2 K – 77 K by means of EPR. The impurity center is attributed to $Mo^{3+}$. No additional line splitting in the EPR spectrum due to the 105 K phase transition has been observed. At 4.2 K the following spin Hamiltonian parameters for this impurity ion were obtained:
$g = 1.9546 \pm 0.0010$ and $A = (32.0 \pm 0.05) \times 10^{-4}$ cm$^{-1}$.

*Introduction*

Octahedrally coordinated $d^3$ impurity ions have been thoroughly studied in $SrTiO_3$ and other ionic crystals.[1] For instance, above the 105 K phase transition the EPR spectrum of the $SrTiO_3:Cr^{3+}$ ($3d^3$) system shows an isotropic line with $g = 1.978$ and 4 resolved hyperfine lines due to the $^{53}Cr$ ($I = 3/2$) isotope. The ionic radius of $Cr^{3+}$ (0.63 Å) is practically the same as that of the $Ti^{4+}$ ion (0.64 Å), which it replaces, so the environment remains cubic. In contrast to this the $Mn^{4+}$ ion ($3d^3$) ion sees an additional field above 105 K of tetragonal symmetry because the radius of $Mn^{4+}$ (0.62 Å) is slightly smaller than that of $Ti^{4+}$ and as a consequence the $Mn^{4+}$ ion sits off-centered. At 77 K, i.e. well below the structural phase transition, one sees in the EPR spectrum of $Cr^{3+}$ additional splittings because the phase transition introduces tetragonal domains and these in turn give rise to a zero field splitting in the spin Hamiltonian for the $S = 3/2$ system.

Here we present results of a new photochromic $d^3$ impurity system in $SrTiO_3$.[2] We discuss the EPR spectra which are characteristic of a $Mo^{3+}$($4d^3$) impurity ion in an almost cubic field.

*Experimental*

Single crystals of $SrTiO_3$ doped with Fe in a concentration of 18±4 ppm and Mo in a concentration of 2±1 ppm were purchased from Semi-Elements Inc. and National Lead Co. The size of the crystals was typically 2×2×2 mm$^3$. EPR measurements were made at X-band (9.2 GHz) utilizing 100 kHz modulation and K-band (19.469 GHz). In the X-band measurements the sample was mounted in an optical transmission cavity. The sample was irradiated with light from a Philips SP 500 W high-pressure mercury arc lamp filtered by a 396 nm filter. The set up at K-band was as follows: A concave mirror collector, with a diameter of 140 mm, focuses the light filtered with a 396 nm filter with 0.9 nm half width from a Xenon arc lamp back onto the entrance of a Suprasil wideband light pipe to which the sample is mounted at the other conical end. The measurements were made in the temperature range of 4.2 K – 77 K.

*Results*

After three minutes illumination at 4.2 K a central line with 6 hyperfine lines could be observed in K-band (Fig. 1). The central line remained visible after warming up to 77 K. Also the following light induced changes of the EPR spectra could be observed: signals due to $Fe^{3+}$ [3], $Fe^{3+}-V_O$ [4] and $Fe^{4+}-V_O$ [5] disappeared and those due to the cubic and non-cubic $Fe^{5+}$ impurities[6, 7] were created. In addition the dynamic Jahn-Teller (JT) $Mo^{5+}$ [8] was observed. In the temperature range of 4.2 K – 77 K the angular dependence of the EPR spectrum with the magnetic field $\vec{H}$ in the (001)-plane shows that both the g- and hyperfine values are isotropic. The spin Hamiltonian parameters corresponding to the spectrum of Fig. 1 are given by:

$g = 1.9545 \pm 0.0010$ and $A = (32.9 \pm 0.05) \times 10^{-4}$ cm$^{-1}$.

Above the structural phase transition of SrTiO$_3$ (T > 105 K) no spectra of the center could be detected, probably due to thermal bleaching back into the parent center.

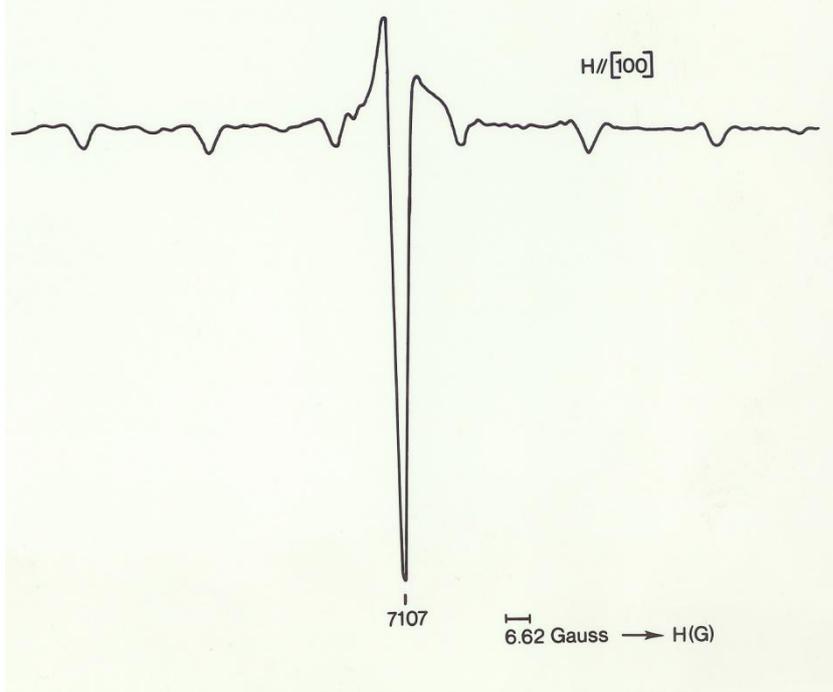

Fig. 1. EPR spectrum of the Mo$^{3+}$ ion in SrTiO$_3$ with $\vec{H} \parallel [100]$ at 19.4 GHz, T = 4.2 K.

*Discussion*

The EPR results show that the impurity center has an effective spin of $S = \frac{1}{2}$ and hyperfine interactions with $I = 5/2$ occur. In fact we attribute the new center to an $S = 3/2$ spin system. Molybdenum has two isotopes, with a natural abundance of 15.8% for $^{95}$Mo ($I = 5/2$) and 9.6% for $^{97}$Mo ($I = 5/2$), which differ in magnetic moment by only 2% (-0.9099 and -0.9290). For the $S = 3/2$ Mo center one central line due to the even isotopes of Mo, which have zero nuclear spin, and six hyperfine lines, which are due to the odd isotopes, are expected. The ratio of the integral intensity for the central line and the sum of the hyperfine lines is 22:7. This fits well with the natural abundance of the molybdenum isotope (Mo$^{even}$ = 75%; Mo$^{odd}$ = 25%).

The spin Hamiltonian describing the energy levels of the Mo$^{3+}$ ($S = 3/2$) ion can be written as follows:[9, 10]

$$\mathcal{H} = \mu_B H . \bar{g} . S + S . \bar{D} . S + S . \bar{A} . I$$

Above the structural phase transition Mo$^{3+}$ is at a cubic site ($D = 0$) and one central isotropic and six hyperfine lines can be observed. Below the structural phase transition a small tetragonal component is present and the above mentioned equation takes the form:

$$\mathcal{H} = g_\parallel \mu_B H_z S_z + g_\perp H_x S_x + H_y S_y] + D[S_z^2 - \frac{5}{4}]$$

In principal, three EPR transitions are possible, the resonance fields being:

$$\mathcal{H} = \frac{h\nu}{g\mu_B} - \frac{D}{2g\mu_B} \frac{3g_\parallel^2 \cos^2\theta - g^2}{g^2}(2M_S - 1)$$

where $g^2 = g_\parallel^2 \cos^2\theta + g_\perp^2 \sin^2\theta$.

$\theta$ is the angle between the applied magnetic field and the tetragonal axis of the $S = 3/2$ center and $M_S$ takes the values ±3/2 and ±1/2.

The separation of the fine structure lines in fist order is given as,

$$\Delta H = \left| \frac{D}{g\mu_B} \frac{3g_\parallel^2 \cos^2\theta - g^2}{g^2} \right|$$

.

In case of an isotropic g-value this becomes,

$$\Delta H = \left| \frac{D}{g\mu_B}(3\cos^2\theta - 1) \right|$$

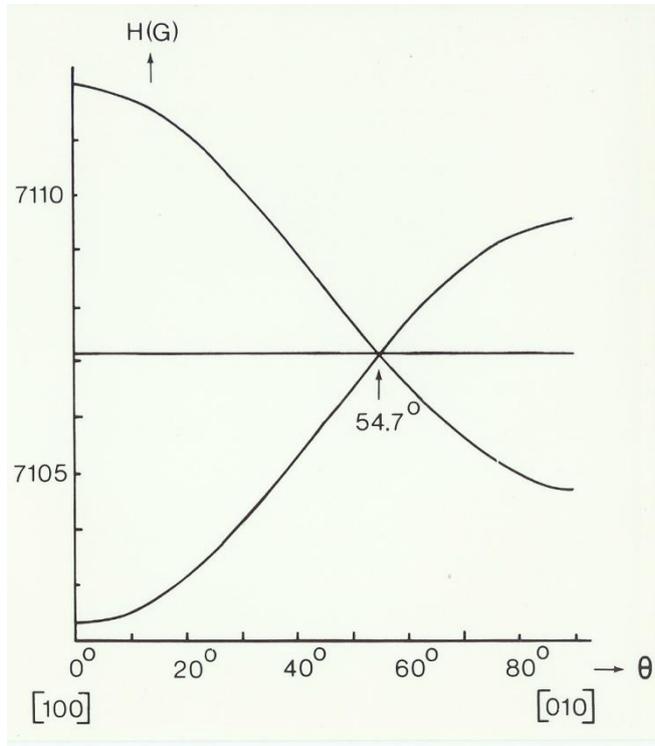

Fig. 2. Angular dependence of the fine structure spectrum of an S = 3/2 ion with g = 1.9546 at 19.456 GHz and a small tetragonal distortion of $|D| = 2.2 \times 10^{-4}$ cm$^{-1}$. The linewidth is smaller than the fine structure splitting.

In Fig. 2 we depict the calculated angular behavior of the fine structure in first order. Below 105 K the crystal is composed of tetragonal domains with their axes in [100], [010] and [001] directions respectively. For $\vec{H}$ parallel to a [100] direction, one obtains three fine structure lines with a separation of 2D from those ions with their main axes parallel to $\vec{H}$. Those $Mo^{3+}$ sites in domains with main axes perpendicular to $\vec{H}$, i.e. parallel to [010] and [001], also yield three fine structure lines but with a separation of D. So the central lines consist in all of five lines separated by D. A further splitting is expected for those $Mo^{3+}$ ions which have a nuclear spin. For these isotopes the spectrum consists of $2I + 1$ equally spaced lines (considering only first-order effects), with resonance fields given by,

$$H = \frac{h\nu}{g\mu_B} - \frac{A}{g\mu_B}M_I$$

The spacing varies with the angle $\theta$ as, $\Delta H = \frac{A}{g\mu_B}$, with $Ag = \left[A_\parallel^2 g_\parallel^2 \cos^2\theta + A_\perp^2 g_\perp^2 \sin^2\theta\right]^{\frac{1}{2}}$

Since the g-factor and hyperfine interactions are isotropic it is concluded that the ground state is an orbital singlet. In second-order perturbation theory the g-factor is expressed as,

$$g = g_S - \frac{8\lambda}{\Delta}$$

where $g_S$ is the free electron spin value, $\lambda$ is the spin-orbit coupling constant and $\Delta$ is the energy difference between ground and first excited state of the same symmetry.[9] For the spin-orbit coupling constant we take the value measured for $Mo^{3+}$ in other systems,[9, 11] i.e. $\lambda = 800$ cm$^{-1}$. Experimentally, we obtain $g = 1.9546$ and from the last equation we obtain for $\Delta \approx 16,770$ cm$^{-1}$ ($\approx 2.08$ eV).

As already remarked above, in our experiments no fine structure splittings were found. From the linewidth of the central line when $\vec{H}$ is parallel to [100], we can estimate an upper limit value for the zero-field splitting 2D in the low temperature phase. The linewidth equals 9.6 Gauss at 4.2 K and thus $\left|\frac{4D}{g\mu_B}\right| \leq 9.6$ Gauss, which gives an *upper* limit of $|D| = 2.2\times10^{-4}$ cm$^{-1}$ at T = 4.2 K.

The g- and A-values of $Mo^{3+}$ in $SrTiO_3$ (at 4.2 K), corundum[12] (at 77 K) and rutile[13] (at 77 K) crystals are given in the table. The somewhat smaller values for the hyperfine constants in $SrTiO_3$ show that there is more covalent bonding than in corundum and rutile. A similar effect was also found in the $SrTiO_3$:$Cr^{3+}$ system.[1] The order of the hyperfine constant for $Mo^{3+}$ compares well with that for $Mo^{5+}$ ($A_\parallel = 39\times10^{-4}$ cm$^{-1}$).

In earlier optical and EPR experiments the photochromic dynamic JT ion $Mo^{5+}$ at 4.2 K was found.[8, 14, 15] The presence of $Mo^{5+}$ at 4.2 K was confirmed in our experiments. Evidently after irradiation, the molybdenum is simultaneously present as 3+ and 5+. No signals of the EPR silent $Mo^{4+}$ could be detected in our experiments.

|   | SrTiO$_3$ | Al$_2$O$_3$ | TiO$_2$ |
|---|---|---|---|
| $g$ | 1.9546 | | |
| $A$ | 32.854×10$^{-4}$ cm$^{-1}$ | | |
| $g_\parallel$ | | 1.968 | |
| $g_\perp$ | | 1.968 | |
| $A_\parallel$ | | 119 MHz | |
| $A_\perp$ | | 121 MHz | |
| $A_{av}$ | | 40×10$^{-4}$ cm$^{-1}$ | |
| $g_z$ | | | 1.94 |
| $g_{[\bar{1}10]}$ | | | 1.97 |
| $g_{[110]}$ | | | 1.95 |
| $A_z$ | | | 38.3×10$^{-4}$ cm$^{-1}$ |
| $A_{[\bar{1}10]}$ | | | 35.0×10$^{-4}$ cm$^{-1}$ |
| $A_{[110]}$ | | | 35.6×10$^{-4}$ cm$^{-1}$ |